\documentclass[pra,aps,twocolumn,showpacs,nofootinbib]{revtex4}
\usepackage{amsfonts,amsmath,bm,bbm, amsthm,amssymb}
\usepackage{enumitem}
\usepackage[OT4]{fontenc}
\usepackage[usenames,dvipsnames]{color}
\usepackage{graphicx}
\usepackage{epstopdf}
\usepackage{verbatim}
\usepackage{multirow}

\usepackage{hyperref}

\hypersetup{
   colorlinks=true,         
   linkcolor=blue,          
   citecolor=blue,        
   urlcolor=black          
}

\newcommand{\ket}[1]{\left| #1 \right>} 
\newcommand{\bra}[1]{\left< #1 \right|} 
\newcommand{\braket}[2]{\left< #1 \vphantom{#2} \right|
 \left. #2 \vphantom{#1} \right>} 

\newcommand{\ketbra}[2]{\ensuremath{\left|#1\right\rangle\!\left\langle#2\right|}}
\renewcommand{\v}[1]{\ensuremath{\boldsymbol{\mathbf{#1}}}}

\newcommand{\be}{\begin{equation}}
\newcommand{\ee}{\end{equation}}

\newcommand{\tr}[1]{\mathrm{Tr}\left[ #1 \right]}

\theoremstyle{plain}

\theoremstyle{definition}

\theoremstyle{remark}

\def\>{\rangle}
\def\<{\langle}
\def\E{ {\cal E} }
\def\H{ {\cal H} }

\def\H{ {\cal H} }

\def\D{ {\cal D} }

\def\I{ \mathbbm{1} }

\begin{document}

\title{Description of quantum coherence in thermodynamic processes requires constraints beyond free energy}
\author{Matteo Lostaglio}
\email{lostaglio@gmail.com}
\author{David Jennings}
\author{Terry Rudolph}
\affiliation{Department of Physics, Imperial College London, London SW7 2AZ, United Kingdom}

\begin{abstract}
Recent studies have developed fundamental limitations on nanoscale thermodynamics, in terms of a set of independent free energy relations. Here we show that free energy relations cannot properly describe quantum coherence in thermodynamic processes. By casting time-asymmetry as a quantifiable, fundamental resource of a quantum state we arrive at an additional, independent set of thermodynamic constraints that naturally extend the existing ones. These asymmetry relations reveal that the traditional Szilard engine argument does not extend automatically to quantum coherences, but instead only relational coherences in a multipartite scenario can contribute to thermodynamic work. We find that coherence transformations are always irreversible. Our results also reveal additional structural parallels between thermodynamics and the theory of entanglement.

\end{abstract}

\maketitle

We are increasingly able to probe and manipulate the physics of micro and nano-scale systems. This has led to the explosion of work in the field of nanotechnology, with a myriad of applications to areas in industry, information technology, medicine and energy technologies. 
With operating scales between $1-10^2$ nm, there has been remarkable progress in the development of molecular information ratchets, molecular motors, optical thermal ratchets and artificial bipedal nanowalkers inspired by naturally occurring biomolecular walkers \cite{collin2005verification, serreli2006molecular, toyabe2010experimental,alemany2010fluctuations, cheng2010bipedal}. There is also increasing evidence for the role of quantum effects within biological systems \cite{lloyd2011quantum,lambert2013quantum,gauger2011sustained}.

Towards the lower-end of the nanoscale, quantum mechanical effects such as quantum coherence and entanglement increasingly make their presence felt. Electrical conductance of molecular-scale components no longer obey Kirchhoff's laws and phase coherence can provide both destructive as well as constructive interference effects on electrical transport \cite{vazquez2012probing}.
Such coherence has been shown to play important roles in thermal to electrical power conversion, heat dissipation in atomic-scale junctions and the engineering toolkit of quantum dots \cite{karlstrom2011increasing}. Conversely, dissipative quantum thermodynamics offers the possibility of on-demand generation of quantum information resources essential for future quantum technologies (communication, encryption, metrology and computing) \cite{lin2013dissipative}. Within quantum information science the question of thermodynamically robust quantum memories, and thermodynamic constraints on quantum computation are still only partially understood and provide deep questions in the overlap between thermodynamics and quantum theory \cite{bravyi2011quantum, pachos2012introduction}. In a similar way, the phenomenon of thermality due to entanglement and the thermodynamics of area laws reveal deep connections between thermodynamics and the theory of entanglement \cite{popescu2005foundations, brandao2008entanglement}. 

The physics of these remarkable small-scale systems, displaying coherence or entanglement, constitute extreme quantum regimes. As such, a crucial question is: to what degree do traditional thermodynamic formulations and techniques encapsulate this regime? This is a broad, foundational question about thermodynamics. It is increasingly apparent that the traditional entropic formulation that emerges as an essentially unique description of the irreversibility of classical, macroscopic systems, will only place necessary, but not sufficient, constraints on the physics of small-scale systems manifesting coherence or quantum correlations.

The textbook treatments of classical, macroscopic equilibrium thermodynamics are typically based on notions such as Carnot cycles, with the entropy function generically defined via an integral in terms of heat flow \cite{fermi1956thermodynamics}. This thermodynamic entropy function is then assumed (but often not proved) to completely describe the irreversible constraints on the system at hand. Alternative approaches follow a statistical mechanical treatment of the system based on underlying microstates, and provide an explanation of the thermodynamics in terms of microscopic degrees of freedom.

However, more rigorous derivations of the entropic form of the second law exist, such as by Carath\'eodory \cite{caratheodory1925uber}, Giles \cite{giles1965mathematical} and more recently by Lieb and Yngvason \cite{lieb1999physics}. Of central importance is the partial order of thermodynamic states, from which an entropy function can then be derived in a rigorous manner. The existence of an essentially unique entropic form of the second law is found to be equivalent to assumptions that fail to hold in small-scale systems or high correlation quantum environments. For example, a scaling hypothesis is required, which is no longer valid for small systems. In addition a ``Comparison Hypothesis'' \cite{giles1965mathematical, lieb1999physics} is required to hold (or derived from other axioms), which in itself makes a highly non-trivial assumption on the structure of the thermodynamic partial order. Outside of the macroscopic classical regime,  quantum systems will generically possess coherence or entanglement, and the ordering of states typically displays a much richer structure \cite{horodecki2009quantum}. 

A unique additive entropic function implies that such assumptions must hold \cite{lieb1999physics}. Therefore their inapplicability in the quantum realm means that no single entropic function can suffice. To fully describe the thermodynamic directionality of nanoscale, non-equilibrium systems, more than one entropy function is required. The results of \cite{brandao2013second} provide a clean characterization of non-asymptotic, thermodynamic inter-conversions of quantum states with zero coherence between energy eigenspaces. The necessary and sufficient conditions for such state inter-conversions are in terms of a set of entropic free energy functions (here denoted $\Delta F_\alpha \le 0$).  The present work goes beyond these conditions, showing that even these fail to be sufficient for thermodynamic transformations involving non-zero quantum coherence.

Exploiting recent results in asymmetry theory \cite{marvian2014extending, marvian2013asymmetry}, we show that thermodynamics can be viewed as being determined by at least two independent resources: the first is quantified by known free energies and measures how far a state is from being thermal; the second, a missing ingredient of previous treatments, measures how much a quantum state breaks time-translation invariance, i.e. the degree of coherence in the system.
This removes the ``zero coherence'' assumption made in numerous recent works, e.g. \cite{egloff2012laws, horodecki2013fundamental, brandao2013second, aberg2013truly}. This shift in perspective allows us to extend the free energy relations to a parallel set of thermodynamic constraints for quantum coherence, which take the form $\Delta A_\alpha \le 0$, where $A_{\alpha}$ are measures of time-translation asymmetry. These constraints characterize the tendency of any quantum system to ``equilibrate'' towards a time-symmetric state.
The new laws, irrelevant for a system composed of many, uncorrelated bodies, become essential for the thermodynamics of small/correlated quantum systems. As an application we show that in certain regimes the free energy splits into two components, one measuring the amount of classical free energy and the other measuring the quantum contribution coming from coherence. We show that coherence is not directly distillable as work, but does admit activation as a relational degree of freedom. We uncover a second form of fundamental irreversibility that parallels the one stressed in \cite{horodecki2013fundamental} but involves coherence transformations. Finally, we shed light on new connections between thermodynamics and entanglement theory.

\section{Results}

\subsection{Free energy Second Laws}

\label{sec:inadequacy}

The approach most suited to our needs in this work is the one followed in \cite{janzing2000thermodynamic, brandao2011resource, horodecki2013fundamental, brandao2013second}, which has emerged from the theory of entanglement \cite{horodecki2009quantum}.
Thermodynamic transformations (also called thermal operations) are defined as the set of all energy-preserving interactions between an arbitrary quantum system and a Gibbsian bath at a fixed temperature (See methods).

One can allow additional, auxiliary systems to be used catalytically and consider thermodynamic transformations $\rho\otimes \chi_{\rm aux} \rightarrow \sigma \otimes \chi_{\rm aux}$, where an auxiliary system begins and ends in the same state $\chi_{\rm aux}$, yet enables the otherwise forbidden thermodynamic transformation $\rho \rightarrow \sigma$. For this broad setting, it was recently proven \cite{brandao2013second} that a continuum of quantum second laws govern the allowed thermodynamic transformations. Specifically the generalized free energies given by $F_{\alpha}(\rho) =kT  S_{\alpha}(\rho||\gamma) - kT \log Z_H$, $Z_H = \mathrm{Tr}[ e^{-\beta H}]$, must all decrease:
\be
\label{freeensecondlaws}
\Delta F_{\alpha}\leq 0, \quad \forall \alpha \geq 0.
\ee
Here $\gamma$ is the thermal state of the system with Hamiltonian $H$, $\gamma = e^{-\beta H}/ Z_H$, $\beta = (kT)^{-1}$ and $S_{\alpha}$ (sometimes denoted $D_{\alpha}$) are information-theoretic generalizations of the standard relative entropy, called $\alpha$-R\'enyi divergences \cite{renyi1961measures} (See methods).
For $\alpha \rightarrow 1$, $S_{\alpha}(\rho||\sigma)$ is simply the quantum relative entropy \cite{muller2013quantum} and the constraints of Eq.~\eqref{freeensecondlaws} reduce to $\Delta F \leq 0$, where $F(\rho) = \mathrm{Tr}[\rho H] - kT S(\rho)$. When applied to isothermal transformations between equilibrium states these conditions reproduce the traditional bound on work extraction \cite{horodecki2013fundamental, brandao2013second} (See methods). However these conditions turn out to be also sufficient for characterizing the states accessible through thermodynamic transformations with the aid of a catalyst, when no coherence is present. For a system of many, uncorrelated particles only $\alpha \rightarrow 1$ matters, so that the family of second laws collapse to the traditional constraint of non-increasing free energy \cite{brandao2011resource, brandao2013second}.

Previous work considered either an asymptotic scenario or assumed the states to be block-diagonal in the eigenbasis of the Hamiltonian. Both these assumptions are insensitive to the role of coherence. Indeed the free energy relations are no longer sufficient for the single-shot thermodynamics of correlated and coherent quantum systems. As we shall see now, additional conditions are required due to the breakdown of time-translation invariance.  

\subsection{Beyond conservation laws}
\label{newsymmetry}

The idea of symmetry is powerful and wide-reaching, and finds countless applications across physics. However, recent work has brought the concept of asymmetry to the fore, and shown it to be a valuable, consumable resource \cite{bartlett2007reference, gour2008resource, spekkens2009relative, marvian2013asymmetry, marvian2013modes, ahmadi2012way}. 
An evolution is said to be symmetric if it commutes with the action of a symmetry group, i.e. it does not matter if the symmetry transformation is applied before or after the dynamics takes place. Similarly, a state is symmetric if it is invariant under symmetry transformations and asymmetric otherwise (See methods). Asymmetric states, in analogy with entangled states, constitute a resource that makes possible transformations otherwise impossible under the constraint of a symmetry group.

It has been found that symmetry constraints for closed system dynamics of pure quantum states (not mixed) are encoded by the conservation of all moments of the generators of the symmetry transformations. However, this is not the case for open quantum system dynamics, or for mixed quantum states, and asymmetry monotones, i.e. functions that do not increase under symmetric evolution \cite{gour2008resource}, can impose further, non-trivial constraints on the dynamics \cite{marvian2014extending} (See methods for a brief discussion of the connections between the present approach and fluctuation theorems).

\subsection{Time-asymmetry and Thermodynamics}

Noether's theorem tells us that if a system has time-translation invariance then its energy is conserved. However, in general thermodynamic scenarios we have no time-translation invariance, either for the thermodynamic process on the system or for the quantum state of the system. The thermodynamics of a system generally involves irreversible dynamics and mixed quantum states out of equilibrium, and heat can flow into and out of the thermal reservoir. 

One might therefore think that the unitary group generated by the free Hamiltonian $H$ of the system should not play any particular role. However this is not the case, and from a perspective of asymmetry we find that:

\emph{Theorem 1:}
The set of thermal operations on a quantum system is a strict subset of the set of symmetric quantum operations with respect to time-translations.

\begin{figure}[ht]
\centering
\includegraphics[width=0.45\textwidth]{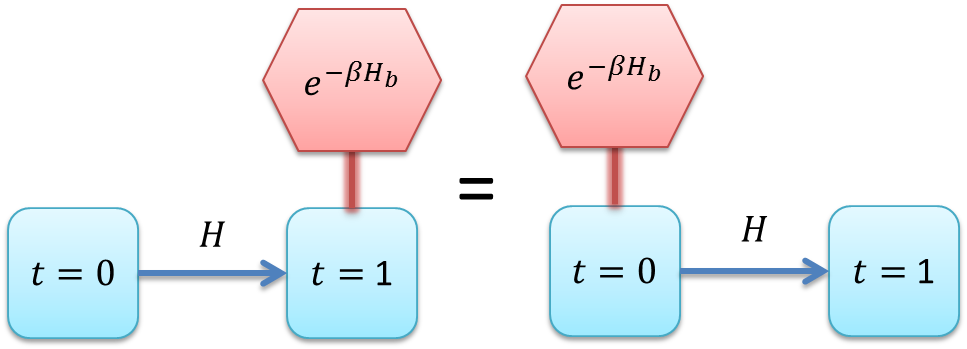}
\caption{\textbf{Time-translation symmetry.} Connecting a thermal bath, with Hamiltonian $H_{\mathrm{b}}$, to a quantum state before or after free time evolution does not make any difference to the resultant state. This simple symmetry implies laws that constrain the approach of a state to time-translation invariance.}
\label{fig:symmetricoperations}
\end{figure}
The proof of this is provided in the Methods. See also Fig.~\ref{fig:symmetricoperations}. The implication of this result is that no thermodynamic process can generate additional time-translation asymmetry in the quantum system. A general picture emerges, where thermodynamics is governed by distinct abstract resources. The ``thermodynamic purity'' resource component, $p$, quantifies how ordered the state of the system is in the presence of a thermal bath, and its evolution is constrained by a set of free energy differences \cite{brandao2013second} (see Methods). If no quantum coherence is present then consideration of $p$ suffices, however more generally quantum thermodynamics is governed by the interplay of at least two fundamental resources, denoted by $(p,a)$. Free energy relations quantify the former, while asymmetry theory provides the tools to quantify the latter.

\subsection{Coherence Second Laws}
\label{sec:coherencesecondlaw}

We now present thermodynamic constraints that go beyond free energy relations. In particular we find that the core measures, used to define the generalized free energy relations \cite{brandao2013second}, can be extended in a natural way that provides asymmetry measures. We introduce the following:

\emph{Definition 1}: for any $\alpha \geq 0$ the free coherence of a state $\rho$ with respect to a Hamiltonian $H$ is
\be
\nonumber
A_{\alpha}(\rho):=S_{\alpha}(\rho || \mathcal{D}_{H}(\rho)),
\ee
where $\mathcal{D}_{H}$ is the operation that removes all coherence between energy eigenspaces. $S_{\alpha}$ are the quantum R\'enyi divergences as defined in the methods.

\begin{figure}[h]
\begin{center}
\includegraphics[width=0.45\textwidth]{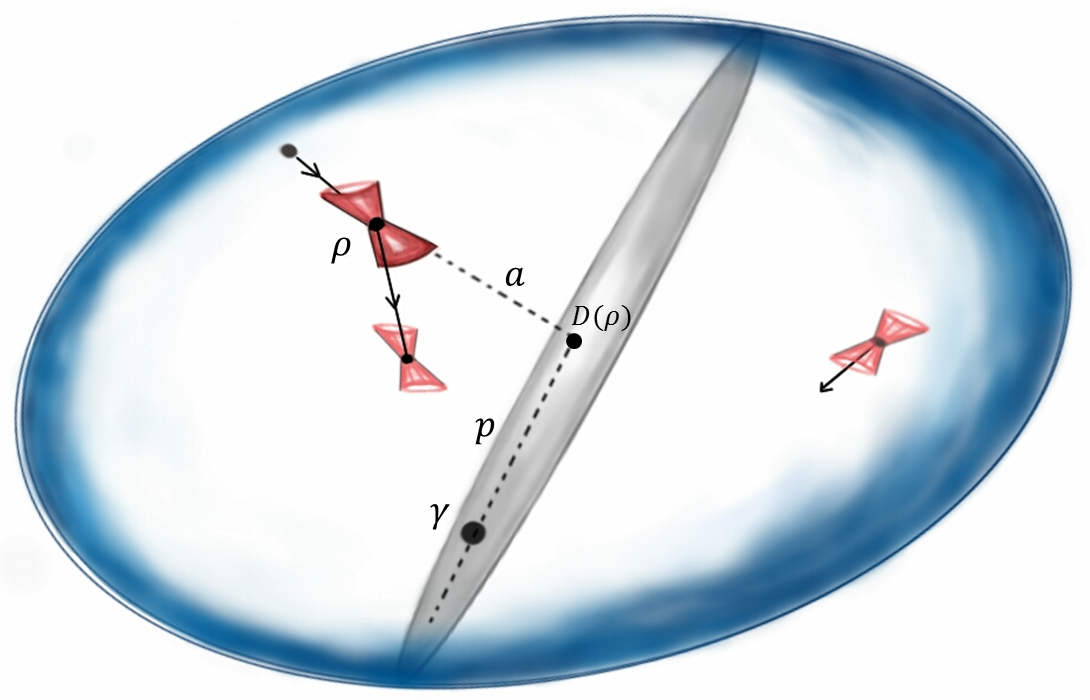}
\end{center}
\caption{\textbf{Quantum Thermodynamics as the combination of asymmetry and thermodynamic purity.} The blue blob is the convex set of all quantum states. To any state $\rho$ we can associate a ``thermal cone'' (in red), the convex set of states thermally accessible from it. Any state $\rho$ contributes in terms of thermodynamic purity $p$, which corresponds to the deviation of $\mathcal{D}_H(\rho)$ from the thermal state $\gamma$ -- as measured by $\{F_\alpha \}$ -- and asymmetry $a$, which corresponds to the deviation of $\rho$ from the manifold of time-symmetric states (the grey region) -- as measured by $\{A_\alpha\}$.}
\label{fig:blob}
\end{figure}

In the same way in which free energies measure ``how far'' a state is from being thermal,  free coherences measure ``how far'' a state is from being incoherent in energy, i.e. time-translation invariant (see Fig.~\ref{fig:blob}). For $\alpha \rightarrow 1$, we have $A_{1}(\rho) \equiv A(\rho)$ which is the asymmetry measure introduced in \cite{aberg2006superposition, spekkens2009relative, plenio2013coherence}. With these definitions on board, and from Theorem~1, we immediately have the following result: 

\emph{Theorem 2}: \label{coherencesecondlaw}
For all $\alpha \ge 0$ we necessarily have $\Delta A_\alpha \le 0$ for any thermal operation.

These laws characterize the depletion of coherence and the tendency to equilibrate onto the manifold of time-translation invariant states. In particular, they also hold for catalytic thermal operations where the catalyst is block-diagonal in the energy eigenbasis and can be extended to time-dependent Hamiltonians (See methods). Importantly,  these provide constraints that are independent of any free energy relations.

The free energy for $\alpha \rightarrow 1$, the relevant measure of average work yield \cite{skrzypczyk2013extracting}, naturally splits into a classical and a quantum contribution
\be
\label{splitting}
F(\rho) = F_{\mathrm{c}}(\rho) + kT A(\rho)
\ee
where $A(\rho) = S(\rho || D_{H}(\rho))$ measures the amount of coherence in the system and $F_{\mathrm{c}} (\rho) = F(\mathcal{D}_H(\rho))$ is the classical free energy. These results, together with the existing free energy relations, allow us to say that for $\alpha \rightarrow 1$ the classical and quantum contributions to the quantum free energy must independently decrease under any thermodynamic process. Notice that a similar result, although differently interpreted, was found in the context of quantum reference frames \cite{janzing2006quantum}.

\subsection{The incompleteness of existing second laws}
\label{inadequacy}

We now establish that the above asymmetry relations are both independent of the free energy relations, and provide additional non-trivial constraints that must be obeyed in any thermodynamic process $\rho~ \rightarrow~\sigma$.
 
To this end, it suffices to consider a qubit system with Hamiltonian $H = \ketbra{1}{1}$, and choose an initial state $\rho = \ketbra{1}{1}$, together with the target final state
\be
\nonumber
\sigma = (1-\epsilon) \mathbb{\gamma}+\epsilon \ketbra{+}{+}.
\ee
Since $S_\alpha$ is monotonically decreasing in $\alpha$, it suffices to choose $\epsilon >0$ sufficiently small so that $S_\infty (\sigma ||\gamma)  \le S_0 (\rho || \gamma)$ to ensure all of the free energy conditions are obeyed. However, since the initial state is a symmetric state, and $A_\alpha(\sigma)>~0$ for any $\epsilon >0$, it follows that such a transformation is impossible to achieve via a thermodynamic transformation. Thus the free energy relations are necessarily incomplete.

Another way of seeing that the free energy relations only provide an incomplete description of thermodynamics is through the notion of work.
Specifically work is taken to be an ordered state of elevated energy. This idealised ``work bit'' is a two-level system with Hamiltonian $H_w = w \ket{w}\bra{w}$ \cite{horodecki2013fundamental}. In its simplest form, it can be thought of as a perfectly controlled atom that gets excited (de-excited) when energy is extracted from (pumped into) a quantum system through a thermodynamic operation, e.g.
\be
\label{transform}
\rho \otimes \ketbra{w}{w} \rightarrow  \sigma \otimes \ketbra{0}{0} .
\ee
Given any two states, $\rho$ and $\sigma$, one can readily show there exists a $w>0$ such that for all $\alpha \geq0$ the free energy conditions $\Delta F_{\alpha} \leq 0$ are satisfied by Eq.~\eqref{transform}. Thus, adding enough work, any state transformation is possible (at least catalytically) between block-diagonal states. In this sense, work is a universal resource classically. However it is easy to see that Theorem~2 implies that for the transformation of Eq.~\eqref{transform} to be possible we need $A_{\alpha}(\rho) \geq A_{\alpha}(\sigma)$, for all $\alpha \geq 0$. In quantum thermodynamics, both the energetic and the coherent properties must be considered together.

\subsection{Emergence of classicality}
\label{classicality}

The constraints of Theorem 2 are not only relevant for nanoscale thermodynamics, but also at the macroscopic scale in the presence of correlated quantum systems able to sustain coherence. The regime in which the coherence second laws may be neglected is for systems composed of many, non-interacting bodies. We formalize this question and answer it, by showing that the free coherences per particle in a system of $n$ non-interacting qubits vanish in the $n \rightarrow \infty$ limit:
\be
\label{classicality}
\lim_{n\rightarrow \infty} A_{\alpha} (\rho^{\otimes n})/n = 0, \quad \forall \alpha \geq 0.
\ee
This generalizes the result found in \cite{spekkens2009relative} for $\alpha \rightarrow 1$ and describes an emergent classical scenario in which states become effectively time-symmetric. This is the reason why only the free energy governs the asymptotic behaviour in \cite{brandao2011resource}. In particular the following bound holds (See methods):
\be
\label{bound}
0 \leq A_{\alpha} (\rho^{\otimes n})  \leq \log (n+1).
\ee
We will use Eq.~\eqref{bound} shortly to study work extraction at the classical-quantum boundary.

\subsection{Quantum Szilard}
\label{workextraction}

The notions of work and heat are the primary concerns of thermodynamics, and with the advent of nanoscale technologies it has been necessary to revisit these time-honoured concepts (see e.g. \cite{allahverdyan2004maximal, alicki2004thermodynamics,dahlsten2011inadequacy, egloff2012laws, alicki2013entanglement, frenzel2014reexamination} and references therein). The analysis of Szilard \cite{maruyama2009colloquium} showed that the information one has about a system has an energetic value in terms of the ordered work one can obtain from a disordered thermal reservoir \cite{toyabe2010experimental}. Specifically, the possession of a single bit of information can be ``burnt'' to obtain $kT \ln 2$ Joules.  More generally, standard thermodynamic arguments imply that given a state $\rho$ of a $d$-dimensional system, in contact with a thermal reservoir at a fixed temperature, we can obtain an amount of work $ W(\rho)=kT (\ln d - S(\rho))$. Previous works \cite{egloff2012laws, horodecki2013fundamental} have shown how to extend this result to deterministic and probabilistic work extraction from single quantum systems with zero coherence across energy eigenspaces. 
However when we encounter quantum states containing coherences we must necessarily take into account the asymmetry constraints. One might think that the work relations extend without alteration, but this is not the case -- quantum coherences cannot be simply converted into ordered energy, and so the standard Szilard result must be modified.

\emph{Theorem 3}: for a general work extraction:
\be
\nonumber
\rho \otimes \ketbra{0}{0} \rightarrow \sigma \otimes \sum_w p(w)\ketbra{w}{w},
\ee
the work distributions $p(w)$ that can be obtained from the states $\rho$ and $\mathcal{D}_H(\rho)$ through time-translation symmetric operations coincide.

See methods for a proof. This phenomenon may be called work-locking, because coherence contributes to the free energy (see Eq.~\eqref{splitting}), but cannot be extracted as work (see also \cite{skrzypczyk2013extracting} and \cite{horodecki2013fundamental}). This also sheds light on the origin of the irreversibility noticed in \cite{horodecki2013fundamental}. On one hand the work necessary to form a state, measured by $F_{\infty}$, is bigger than the work that we can draw from it, given by $F_0$, because $F_0 < F_{\infty}$. This first irreversibility is not an intrinsically quantum phenomenon, as it is a sole consequence of the free energy constraints of \cite{brandao2013second}. Indeed this same irreversibility is present even for diagonal states (probability distributions) undergoing thermal operations (particular stochastic processes), a classical - albeit not deterministic - theory. However, quantum coherence adds another layer of irreversibility, as the work necessary to generate the coherent part of a quantum state cannot be extracted later, due to the fact that thermodynamic operations are time-translation symmetric quantum maps.

In the thermodynamic limit the work-locking phenomenon is undetectable.  From Eqs.~\eqref{splitting} and \eqref{bound} we have 
$F(\rho^{\otimes n}) \approx F\left(\mathcal{D}_{\sum_i H_i} (\rho^{\otimes n})\right)$,
when $n  \gg \log n $. The free energy is effectively classical, and the maximum extractable work per system approaches the quantum free energy.
Eq.~\eqref{bound} provides a bound on the rate of this suppression, and we find for $n$ qubits,
\be
\nonumber
\frac{A(\rho^{\otimes n})}{F(\rho^{\otimes n})} \leq \frac{\log (n+1)}{n \log 2},
\ee 
independently of the temperature. For example, a naive application of this result to the case of $5$ qubits shows that up to $50 \%$ of the free energy could be locked in coherences, whereas this number falls to $1 \%$ for a system of $1000$ qubits.

\subsection{Coherent activation of work}

We have established that one must associate to a state both purity and asymmetry, abstractly denoted $(p, a)$, and have shown that coherences in isolation do not contribute to thermodynamic work. Schematically, if $(p,a) \rightarrow W$ then $(p,0) \rightarrow W$ too. It might appear that quantum coherences have no effect on the work output of a thermodynamic process, but this is not the case. 

In the case of states block-diagonal in energy eigenspaces, any state that cannot be prepared under thermal operations can be converted into mechanical work. In the fully quantum-mechanical setting this is no longer the case. There are states that cannot be prepared through thermal operations from which it is impossible to draw any useful work. These are precisely the states $\rho$ with coherence for which $ \mathcal{D}_{H}(\rho)=\gamma$. An extreme case is the pure state 
\be
\label{cts}
 \ket{\psi_c} =  Z_H^{-1/2}\sum_k e^{-\beta E_k/2} \ket{E_k},
\ee
where $\ket{E_k}$ are the eigenstates of $H$ and $Z_H$ is the partition function of $H$. However, while $(0,a) \rightarrow (W=0)$, it turns out coherence can be activated in the presence of other quantum systems with coherence:
\begin{equation}
\label{eq:activation}
(0, a_1) + (0,a_2) \rightarrow (W \ne 0).
\end{equation}
By $A + B \rightarrow C$ we mean it is possible to transform $A$ and $B$ jointly into $C$, using thermal operations only. One might expect, following Szilard, that any pure state should yield $kT \log d$ of work, but if this pure state has $a \neq 0$ this is impossible. Eq.~\eqref{eq:activation} tells us the only way to get $kT \log d$ is to smuggle in coherent resources. Only if we allow the use of an external source of coherence does this extraction of work become possible \cite{aberg2014catalytic}.

The way in which coherence in a state $\rho$ can be utilized to obtain mechanical work is readily seen from asymmetry theory and the theory of quantum reference frames \cite{bartlett2007reference}. Having shown that thermal operations commute with time-translation, all the results concerning work extraction under the presence of a superselection rule (e.g. \cite{vaccaro2008tradeoff}) can be immediately applied to thermodynamics. If we have \emph{two} quantum systems in states $\rho_1$ and $\rho_2$ respectively, for which $\D_{H_1} (\rho_1) = \gamma$ and $\D_{H_2} (\rho_2) = \gamma$, then individually no mechanical work can be obtained in the presence of a thermal reservoir. However the two systems can instead encode relational coherence that is accessible. Specifically the introduction of the second system gives $\D_{\bar{H}} (\rho_1 \otimes \rho_2) =\sigma_{12} \ne \gamma \otimes \gamma$, where $\bar{H} = H_1 \otimes \I + \I \otimes H_2$. 
This is also why collective actions on multiple copies can extract work in a situation in which operations on single copies would be useless \cite{skrzypczyk2013extracting}.

Alternatively, we can distinguish one of the systems as being the dominant reference. This perspective admits a different physical interpretation. We take the dimension of $\H_2$ to be much larger than $\H_1$, and the state $\rho_2$ to be highly asymmetric compared to $\rho_1$. The function of $\rho_2$ is now to allow the simulation of a non-symmetric operation $\tilde{\E}$ on the first system:
\be
\tilde{\E} (\rho_1) =  \mathrm{Tr}_2 [ \E ( \rho_1 \otimes \rho_2)]
\ee
A catalytic property in the use of the reference has been recently pointed out in \cite{aberg2014catalytic} and shown to be a consequence of the fact that time-translations are an abelian group.

\section{Discussion}
\label{outlook}

In \cite{horodecki2013fundamental, brandao2013second} the authors showed that the work needed to create a state $\rho$ is measured by $F_{\infty}(\rho)$ and the work extractable is given by $F_{0}(\rho)$. This revealed an inherent irreversibility of thermodynamic transformations. We can now show that a similar irreversibility characterizes the thermodynamic processing of coherence. While normally one wishes to distill out ordered energy via a thermodynamic process, we could equally ask to obtain a high degree of coherence in the final output state under the allowed quantum operations. One could wish to obtain a $d$-dimensional uniform superposition of energy states, $|\I (d)\> := d^{-1/2}\sum_k |k\>$. Conversely, we may want to know how much coherence is needed to create a quantum state. If $\sigma_{\mathrm{sym}}$ is some incoherent quantum state, Theorem 2 requires
\begin{eqnarray}
\nonumber
\rho \rightarrow \sigma_{\mathrm{sym}} \otimes \ketbra{\I (d_{\mathrm{out}})}{\I (d_{\mathrm{out}})} & \Rightarrow & \log d_{\mathrm{out}} \leq A_{0}(\rho) \\
\nonumber
\ketbra{\I (d_{\mathrm{in}})}{\I (d_{\mathrm{in}})} \otimes \sigma_{\mathrm{sym}} \rightarrow \rho & \Rightarrow & \log d_{\mathrm{in}} \geq A_{\infty}(\rho),
\end{eqnarray}
which shows that a further, fundamental irreversibility affects coherence processing as at least $A_{\infty}(\rho) - A_0(\rho)$ amount of coherence is lost in a cycle.

{\footnotesize
\begin{table}
\label{thermoentanglement}
\begin{tabular}{ c | c | c |}
 \cline{2-3}                  
   & Quantum & Entanglement \\
& Thermodynamics & Theory \\ 
 \hline  \hline 
 \multicolumn{1}{ |c|  }{Asymptotic conversion} & Rel. entropy \cite{brandao2011resource} & Rel. entropy \cite{brandao2008entanglement}\\ 
\multicolumn{1}{ |c|  }{$\rho^{\otimes n} \rightarrow \sigma^{\otimes m}$} & $F(\rho) = S(\rho||\gamma)$ 
  & $\inf_{\sigma \in \mathcal{S}} S(\rho||\sigma)$  \\
\hline 
\multicolumn{1}{ |c|  }{\multirow{2}{*}{$W \rightarrow (p,0) \rightarrow W' < W$} } & \multirow{2}{*}{Non-cyclicity \cite{horodecki2013fundamental}} & Ent. formation $\neq$ \\
\multicolumn{1}{ |c|  }{} & & Ent. distillation \\
\hline 
\multicolumn{1}{ |c|  }{\multirow{2}{*}{$(p,a) \rightarrow W \leftarrow (p,0)$}} & Work & Bound \\
 \multicolumn{1}{ |c|  }{}& locking & entanglement \cite{horodecki1998mixed}\\
  \hline  
\multicolumn{1}{ |c|  }{\multirow{2}{*}{$(0, a_1) + (0, a_2) \rightarrow W$}} & Coherence & Entanglement \\
 \multicolumn{1}{ |c|  }{} & activation & activation \cite{horodecki1999bound}\\
\hline
\end{tabular}
\caption{{\small \textbf{Structural parallels.} Quantum thermodynamics and entanglement manipulations present many structural parallels. The asymptotic interconversion of states are governed by relative entropy to the Gibbs states $\gamma$ and the relative entropy to the manifold of separable states $\mathcal{S}$, respectively. The work necessary to create a state is bigger than the work extractable from it; this similarly happens with entangled state creation and distillation. There are states that cannot be created under thermal (LOCC) operations from which no work (entanglement) can be extracted, but the resource can be activated.}}
\end{table}
}

Shortly after the present work, results appeared \cite{cwiklinski2014limitations} on the reduction of quantum coherence under thermal maps, including tight bounds for qubits. Going beyond this, the work of \cite{lostaglio2014quantum} applies the framework developed here to obtain both upper and lower bounds on coherence evolution for general quantum systems. In particular it highlights that the structure of the bounds in \cite{cwiklinski2014limitations} is symmetry-based, and that coherence in thermodynamics admits a broader mode-decomposition in terms of spectral analysis.

Our results also shed light on the structural relationships between entanglement theory and thermodynamics \cite{popescu2005foundations,brandao2008entanglement, jennings2010entanglement} (see Table 1). Beyond structural parallels, this work paves the way for an explicit unification of the resource theories presented here, and of the now well-developed theory of entanglement.

The resource-theoretic perspective is just one recent approach to the thermodynamics of quantum systems, however we argue that this framework presents an elegant and compact perspective on quantum thermodynamics in terms of the interconversion and quantification of two abstract properties: thermodynamic purity and time-asymmetry. These seem to be necessary components in any unified framework that seeks to describe coherent processes and generic quantum thermodynamic phenomena with no classical counterpart.

\footnotesize
\section{Methods}

\subsection{Thermal operations} They are all quantum operations $\mathcal{E}$ of the form  \cite{janzing2000thermodynamic, brandao2011resource}:
\be
\label{thermalop}
\mathcal{E}(\rho) = \mathrm{Tr}_2[ U(\rho \otimes \gamma_{\mathrm{b}})U^{\dag}]
\ee
where $\gamma_{\mathrm{b}} = e^{-\beta H_{\mathrm{b}}} / \mathrm{Tr}[e^{-\beta H_{\mathrm{b}}}]$, $\beta = (kT)^{-1}$, $[U, H \otimes \I + \I \otimes H_{\mathrm{b}}]=0$ and $U$ is a joint unitary on system and environment. $H$ is the Hamiltonian of the system and $H_{\mathrm{b}}$ the Hamiltonian of the environment. The more traditional formulation of thermodynamic processes involves time-dependent Hamiltonians. However, as already noted in \cite{horodecki2013fundamental, brandao2013second}, this framework can encompass such scenarios through the inclusion of a clock degree of freedom. No restrictions are imposed on the initial and final state of the system, which are in general far from equilibrium, nor on the properties of bath or its final state. The interaction is required to preserve total energy, so differently from traditional treatments all external sources of energy (e.g. a work source) must be included in the picture and described quantum-mechanically.

\subsection{Quantum R\'enyi divergences} There are two non-commutative extensions of the notion of $\alpha-$R\'enyi divergence \cite{muller2013quantum, wilde2014strong}. They enjoy operational significance in the regimes $\alpha >1$ and $\alpha <1$, respectively, and coincide with the traditional quantum relative entropy for $\alpha \rightarrow 1$ \cite{mosonyi2013quantum}. This suggests to follow \cite{mosonyi2013quantum} and define
\be
\nonumber
S_{\alpha}(\rho||\sigma) =
\begin{cases} 
\frac{1}{\alpha -1}\log \mathrm{Tr}[\rho^{\alpha} \sigma^{1-\alpha}], \; \, \quad \quad \quad \quad \quad \alpha \in [0,1) \\
\frac{1}{\alpha-1}\log \tr{\left(\sigma^{\frac{1-\alpha}{2\alpha}}\rho \sigma^{\frac{1-\alpha}{2\alpha}}\right)^\alpha}, \quad \alpha >1
\end{cases}
\ee 
The limit for $\alpha \rightarrow 1$ is given by $S_1(\rho||\sigma) = \mathrm{Tr}[\rho (\log \rho - \log \sigma)].$

\subsection{Consistency with equilibrium thermodynamics}

When a system undergoes an isothermal transformation from an equilibrium state with respect to Hamiltonian $H_1$ to an equilibrium state with respect to Hamiltonian $H_2$, in absence of work, then
\be
\label{eq:standard}
F(H_1) \geq F(H_2),
\ee
where $F(H)=-kT \log Z_H$ is the thermodynamic free energy and $Z_H$ is the partition function. The above relation is recovered within the present framework, which shows consistency with the traditional account.

In \cite{brandao2013second} it is shown that the necessary and sufficient condition for a transformation to be possible between two incoherent quantum states $\rho$ and $\sigma$ while the Hamiltonian is changed from $H_1$ to $H_2$ is
\be
\label{aaa}
F_{\alpha}(\rho,H_1) \geq F_{\alpha}(\sigma, H_2) \quad  \forall \alpha \geq 0,
\ee
where $F_{\alpha}(\rho,H) = kT S_{\alpha}(\rho|| \gamma_{H}) - kT \log Z_H$, $\gamma_H = e^{-\beta H}/Z_H$ and $Z_H = \mathrm{Tr}[e^{-\beta H}]$. For any $\alpha$, $F_{\alpha}(\gamma_H, H) = -k T \log Z_H$ and therefore if the initial and final states are thermal, $\rho = \gamma_{H_1}$ and $\sigma = \gamma_{H_2}$, all the conditions of Eq.~\eqref{aaa} are equivalent to Eq.~\eqref{eq:standard}, which fully characterizes the transformation $\gamma_{H_1} \rightarrow \gamma_{H_2}$ between equilibrium states. However, under a broader class of non-equilibrium operations, the conditions of Eq.~\eqref{aaa} are necessary and sufficient to characterize thermodynamic transformations between two non-equilibrium quantum states, provided that no coherence is present.

The notion of work in the present approach is given by the notion of a work bit \cite{horodecki2013fundamental}, as explained in Section~\ref{inadequacy}. We can recover another traditional bound by looking at how much work one can extract in the transformation between two equilibrium states:
\be
\gamma_{H_1} \otimes \ketbra{0}{0} \rightarrow \gamma_{H_2} \otimes \ketbra{w}{w}
\ee
Then all Eqs.~\eqref{aaa} collapse to the condition
\be
w \leq F(H_1) - F(H_2)
\ee
as expected from traditional treatments.

\subsection{Symmetric operations} Let $G$ be a Lie group representing a symmetry, and consider a representation of $G$ on a Hilbert space $\H$ given by $U:g \mapsto U_g$, where $g \in G$ and $U_g$ is a unitary on $\H$. 
A quantum operation $\mathcal{E}_{G}:\mathcal{B}(\mathcal{H}) \rightarrow \mathcal{B}(\mathcal{H})$ is called symmetric if  \cite{bartlett2007reference, marvian2013asymmetry, marvian2014extending}: 
\be
\label{eq:symmetricop}
\mathcal{E}_G(U(g) \rho U^{\dag}(g)) =U(g)  \mathcal{E}_G(\rho ) U^{\dag}(g), \quad \forall \rho, \forall g \in G.
\ee
A state is called symmetric if it is invariant under symmetry transformations, $U(g)\rho U^{\dag}(g) = \rho$. An intuitive example is for the $SU(2)$ representation of the rotation group in 3-dimensions. The group action defines rotations of quantum states, and those that are invariant (such as the singlet state on two spins) are rotationally symmetric, while all others are asymmetric.

\subsection{Connection to Fluctuation Theorems}

We can compare our framework with well-established results in non-equilibrium thermodynamics. Specifically, we compare with fluctuation theorem approaches \cite{jarzynski1997nonequilibrium, crooks1999entropy, jarzynski2004classical, jarzynski2011equalities} that supply powerful descriptions of systems far from equilibrium. The approach of such fluctuation theorems has significant limitations that are not present in the approach taken here. Firstly, while fluctuation theorems can be written down for quantum systems, they only capture stochastic effects which are ``effectively classical'' in nature. More specifically, the requirement of destructive measurements on the initial state unavoidably kills any coherence between energy eigenspaces, and unavoidably kills entanglement between systems. Attempts to generalize to positive operator valued measures (POVMs) quickly hit obstacles when it comes to the pairing of time-reversed trajectories.
As such, only a limited set of quantum mechanical features can be currently addressed through fluctuation theorems.

Another issue is the focus on the expectation values of random variables -- for example the moments of work-gain. For small systems the distributions involved can be quite broad and structured and so it is arguably more natural to analyse it in finer terms, such as those developed within ``single-shot'' regimes \cite{dahlsten2011inadequacy, yunger2014unification}.
More significantly, it has been shown recently that even if you knew all the moments $\<\hat{O}^k\>$ of a quantum observable $\hat{O}$, this is \emph{insufficient} to describe the mixed state quantum mechanics of a system in the presence of a conservation law on $\hat{O}$ \cite{marvian2014extending}. In our case, the consequences of energy conservation are not fully captured by energy measurements, the reason being that coherence properties must be also taken into account. Our work is a first step in this direction.

The common feature of these points is that the traditional approaches, when applied to more and more extreme quantum systems, hit against a range of obstacles.
The single-shot thermodynamics that has recently emerged has been shown to be consistent with existing thermodynamics, but nevertheless does not suffer from any of the above points. Indeed since it has been developed from entanglement theory and the theory of quantum information, the framework is ideally suited to describe such phenomena.

\subsection{Proof of Theorem 1}. 
We need to prove (see Eq. \eqref{eq:symmetricop}):
\be
\label{thermalaresymm}
\forall t, \quad \mathcal{E}(e^{-i H t}\rho e^{i H t}) = e^{-i H t} \mathcal{E}(\rho) e^{i H t}.
\ee
For any bath system $\gamma_{\mathrm{b}} \propto e^{-\beta H_{\mathrm{b}}}$. From Eq.~\eqref{thermalop}, Eq.~\eqref{thermalaresymm} follows using $[H_{\mathrm{b}}, \gamma_{\mathrm{b}}]=0$ and $[U, H + H_{\mathrm{b}}]=0$. That these operations form a proper subset is seen from the fact that transforming an energy eigenstate into any other energy eigenstate is a symmetric operation, but not a thermally allowed operation.

\subsection{Thermodynamic purity}

Our main result shows that a fundamental resource for the thermodynamics of coherent quantum states is time-translation asymmetry. Previous work \cite{brandao2011resource} has already identified the ``thermodynamic purity'' $p$ of a quantum state as a resource for thermodynamics. We speak of ``thermodynamic purity'' because, as we shall see, purity in the thermodynamic framework appears within an embedding that takes the Gibbs state to the maximally mixed state. The mapping is effectively the same as that between the canonical and microcanonical ensembles in textbook treatments. While our main results show that thermodynamics is a special resource theory of asymmetry, the ideas that we briefly summarize here show that thermodynamics is a special theory of purity (see Appendix D of \cite{brandao2013second} for details). The need for two sets of second laws arises from this duality.

The problem solved in \cite{brandao2013second} is to give necessary and sufficient conditions for the existence of a stochastic operation $\Lambda_{\mathrm{th}}$ that maps a probability distribution $p$ to $p'$ through the aid of a catalyst and leaves the thermal state unchanged:
\be
\label{eq:thermotrumping}
\Lambda_{\mathrm{th}}(p \otimes q) = p' \otimes q, \quad \Lambda_{\mathrm{th}}(\gamma \otimes \eta_{\mathrm{c}})=\gamma \otimes \eta_{\mathrm{c}},
\ee
where $\gamma$ is the thermal state of the system (for simplicity take initial and final Hamiltonian to coincide). Notice that the catalyst can be taken to have a trivial Hamiltonian \cite{brandao2013second}, so that the thermal state of the catalyst is a uniform distribution, denoted by $\eta_{\mathrm{c}}$. Here $q$ acts as a catalyst and as such must be given back unchanged.  Notice that this approach is limited to quantum states diagonal in the energy basis. An important observation is that operations that leave the thermal state unchanged are equivalent (in terms of interconversion structure) to thermal operations as defined in \cite{janzing2000thermodynamic, brandao2011resource, horodecki2013fundamental} only if we limit ourselves to initial and final states diagonal in energy (see also \cite{faist2014gibbs}).

A problem similar to \eqref{eq:thermotrumping} was solved in \cite{ruch1978mixing} for stochastic maps having the uniform distribution (rather than the thermal distribution) as a fixed point. This was done through an extended notion of majorization, called trumping. Given two probability distributions $p$ and $p'$ we say that $p$ can be trumped into $p'$ if and only if there exists a probability distribution $q$ (the catalyst) such that $p \otimes q$ majorizes $p' \otimes q$. From the Birkhoff-von Neumann theorem this is equivalent to the existence of a stochastic map $\Lambda$ such that 
\be
\label{eq:trumping}
\Lambda(p \otimes q) = p' \otimes q, \quad \Lambda(\eta \otimes \eta_c)=\eta \otimes \eta_c,
\ee
where $\eta$ and $\eta_c$ are uniform distributions. Notice that stochastic maps which leave the maximally mixed state unchanged give rise to an interconversion structure which is essentially the same as the resource theory of purity studied previously \cite{horodecki2003reversible}. A necessary and sufficient condition for Eq. \eqref{eq:trumping} to hold was given in \cite{klimesh2007inequalities}:
\be
\label{nstrumping}
S_{\alpha}(p||\eta) \geq S_{\alpha}(p'||\eta), \quad \forall \alpha
\ee 
Here $S_{\alpha}(\cdot||\eta)$ are relative R\'enyi divergences w.r.t. the uniform distribution $\eta$. They measure how pure (i.e. far from uniform) a distribution is. To exploit these results to solve problem \eqref{eq:thermotrumping} we need a map embedding the thermal state into the uniform distribution. Given integers $\v{d}=\{d_1,...,d_n\}$, $\sum_i d_i =N$, the authors of \cite{brandao2013second} defined an embedding map $\Gamma_{\v{d}}$ as
\be
\Gamma_{\v{d}} (p) = \oplus_i p_i \eta_i,
\ee
where $\eta_i$ is the uniform distribution of dimension $d_i$. $\Gamma_{\v{d}}$ is a map from a space of $n$-dimensional distributions, that could be called canonical space, to a space of $N$-dimensional probability distributions, that could be called microcanonical space. The reason for these names is as follows. Let us assume for simplicity that the thermal distribution $\gamma$ is rational. Then it is easy to see that there exists $\v{d}$:
\be
\Gamma_{\v{d}} (\gamma) = \eta,
\ee
where $\eta$ is the $N$-dimensional uniform distribution.

Crucially, it is possible to show that
\be
\label{freeenergypurity}
S_{\alpha}(p||\gamma)=S_{\alpha}(\Gamma_{\v{d}} (p)||\eta),
\ee
Because $F_\alpha(p)=k T S_{\alpha}(p||\gamma) - kT \log Z_H$, this shows that the free energy and the purity measures are mapped one into the other by $\Gamma_d$ and its inverse $\Gamma_{\v{d}}^*$. For this reason the generalized free energy differences can be considered measures of thermodynamic purity within the thermodynamic setting. As a consequence of Eq.~\eqref{freeenergypurity} and the definition of $F_\alpha$, if all $F_\alpha$ decrease the purity measures in the embedding space will decrease as well:
\be
\nonumber
F_{\alpha}(p) \leq F_{\alpha}(p') \, \, \Leftrightarrow S_{\alpha}(\Gamma_{\v{d}} (p)||\eta) \leq S_{\alpha}(\Gamma_{\v{d}} (p')||\eta)
\ee
Using the necessary and sufficient conditions for trumping (i.e. that Eqs. \eqref{eq:trumping} and \eqref{nstrumping} are equivalent), this implies that the condition of decreasing $F_\alpha$ is equivalent to the existence of a stochastic map $\Lambda$ that preserves $\eta \otimes \eta_c$ and maps $\Gamma_{\v{d}}(p)$ to $\Gamma_{\v{d}}(p')$ through a catalyst $q$. The map $\Lambda_{\mathrm{th}}=(\Gamma_{\v{d}}^*~\otimes ~\mathbb{I})\Lambda(\Gamma_{\v{d}} \otimes \mathbb{I})$ is a stochastic map from the canonical space to itself. As required,
\begin{eqnarray*}
\Lambda_{\mathrm{th}}(p \otimes q) & = &(\Gamma_{\v{d}}^* \otimes \mathbb{I})\Lambda(\Gamma_{\v{d}}(p)\otimes q) =  \\ 
& = & (\Gamma_{\v{d}}^* \otimes \mathbb{I})(\Gamma_{\v{d}}(p')\otimes q) \; \,  =  p' \otimes q.
\end{eqnarray*}
Moreover, $\Lambda_{\mathrm{th}}(\gamma \otimes \eta_c)=\gamma \otimes \eta_c$, so it is thermal. The embedding map shows a duality between a theory of purity in the microcanonical space and thermodynamics in the canonical space. This proves the decreasing of all generalized free energies $F_{\alpha}$ is equivalent to the existence of $\Lambda_{\mathrm{th}}$ and $q$ satisfying Eq.~\eqref{eq:thermotrumping} (with the catalyst having a trivial Hamiltonian). 

The embedding maps carry zero coherence deviation from equilibrium into the consideration of the purity resource theory on a larger space. Hence it is not possible to handle coherence in this construction. As a consequence, it is necessary to go beyond free energy relations to capture the role of quantum coherence in thermodynamics.

\subsection{Beyond free energy constraints}

\emph{Proof of Theorem 2}. By assumption there exists some thermal operation $\E$ such that $\sigma = \mathcal{E}(\rho)$. Since $\mathcal{E}$ is a thermal operation then it is symmetric (Theorem 1). Integrating Eq.~\eqref{thermalaresymm} over $t$ gives \cite{bartlett2007reference}
\be
\label{edcommute}
[\mathcal{E},\mathcal{D}_{H}] = 0.
\ee
Using \eqref{edcommute} and the data processing inequality for quantum R\'enyi divergences \cite{mosonyi2013quantum, lieb2013monotonicity, beigi2013sandwiched, datta2014limit}, we deduce the coherence second laws.

The coherence second laws presented in Theorem 2 hold also for a broader set of operations, which allow the aid of a catalyst block-diagonal in energy, as in \cite{brandao2013second}:

\emph{Definition 2}:
We say that a state $\rho$ in $\mathcal{H}$ is transformed into state $\sigma$ through a catalytic thermal operation
\be
\label{catthermalop}
\rho \stackrel{\mathrm{cat}}{\rightarrow} \sigma,
\ee
if there are a quantum state $\rho_c$ in a Hilbert space $\mathcal{H}_{\mathrm{c}}$ with Hamiltonian $H_c$ and a thermal operation $\mathcal{E}$ on $\mathcal{H} \otimes \mathcal{H}_\mathrm{c}$:
\be
\mathcal{E}(\rho \otimes \rho_\mathrm{c}) = \sigma \otimes \rho_\mathrm{c}.
\ee
 
\emph{Theorem 4}:
Catalytic thermal operations with a block-diagonal catalyst are symmetric operations,  i.e. if $H$ is the system's Hamiltonian and $\mathcal{C}$ is a catalytic thermal operation,
\be
\mathcal{C}(e^{-i H t}\rho e^{i H t}) = e^{-i H t}\mathcal{C}(\rho) e^{i H t}
\ee

\emph{Proof}: A state $\rho$ is sent to $\rho'$ through a catalytic thermal operation with diagonal catalyst if there exists a state $\sigma$, s.t. $[\sigma, H_\mathrm{c}] = 0$, and a thermal operation $\mathcal{E}$: $\mathcal{E}(\rho~\otimes~\sigma) = \rho ' \otimes \sigma$. We show that the quantum map $\mathcal{C}(\rho)=\mathrm{Tr}_2~\mathcal{E}(\rho~\otimes~\sigma) = \rho'$ is symmetric. Define $H_{\mathrm{tot}} = H + H_\mathrm{c} + H_\mathrm{b}$, sum of the Hamiltonians of system, catalyst and bath. Notice that $\sigma =  e^{-i H_\mathrm{c} t}\sigma e^{i H_\mathrm{c} t}$, $\gamma_{\mathrm{b}} =  e^{-i H_{\mathrm{b}} t}\gamma_{\mathrm{b}} e^{i H_{\mathrm{b}} t}$. It follows
\begin{eqnarray*}
\mathcal{C}(e^{-i H t}\rho e^{i H t}) &= & \mathrm{Tr}_2\mathcal{E}(e^{-i H t}\rho e^{i H t} \otimes \sigma) \\
 & = & \mathrm{Tr}_{23} [ U e^{-i H t} \rho e^{i H t} \otimes \sigma \otimes \gamma_{\mathrm{b}}  U^{\dag}] \\
& = & \mathrm{Tr}_{23} [ e^{-i H_{\mathrm{tot}} t} U \rho \otimes \sigma \otimes  \gamma_{\mathrm{b}}  U^{\dag} e^{-i H_{\mathrm{tot}} t}]   \\
& = & e^{-i H t} \mathrm{Tr}_2[\mathcal{E}(\rho \otimes \sigma)] e^{i H t} \\
& = & e^{-i H t}\mathcal{C}(\rho) e^{i H t}.
\end{eqnarray*}

\emph{Theorem 5}: If $[\rho_\mathrm{c}, H_\mathrm{c}]=0$,
\be
\rho \stackrel{\mathrm{cat}}{\rightarrow} \sigma \Rightarrow A_{\alpha}(\sigma) \leq A_{\alpha}(\rho), \; \; \forall \alpha \geq 0.
\ee

\emph{Proof}: Follows from Theorem 4 in the same way in which Theorem 2 follows from Theorem 1.

\subsection{Coherence second laws for time-dependent Hamiltonians}

In many thermodynamic applications, Hamiltonians are time-dependent. As shown in \cite{horodecki2013fundamental, brandao2013second} we can deal with these situations introducing a classical degree of freedom representing a clock system. This classical degree of freedom can be thought of as a switch for changing the Hamiltonian (e.g. the knob tuning a magnetic field). Hence we might interpret the transformation
\be
\rho \otimes \ketbra{0}{0} \rightarrow \sigma \otimes \ketbra{1}{1}
\ee
as the transformation sending $\rho$ with initial Hamiltonian $H_0$ to $\sigma$ with Hamiltonian $H_1$, once we formally define the Hamiltonian
\be
\label{eq:changinghamiltonian}
H= H_0 \otimes \ketbra{0}{0} + H_1 \otimes \ketbra{1}{1}
\ee
which can be interpreted as a Hamiltonian $H_0$ that changes into $H_1$ as the switch goes from $0$ to $1$. Then Theorem 2 admits the following natural extension:

\emph{Theorem 6}: for all $\alpha \ge 0$ we necessarily have $\Delta A_\alpha \le 0$ for any thermal operation between $\rho$ and $\sigma$ in which the Hamiltonian is switched from $H_0$ to $H_1$. Here we defined
\be
\label{thm:secondlawschanginghamiltonians}
\Delta A_{\alpha} = A_{\alpha}(\sigma,H_1) - A_{\alpha}(\rho,H_0),
\ee
where $A_{\alpha}(\cdot,H_i) = S_{\alpha}(\cdot || \mathcal{D}_{H_i}(\cdot))$.

\emph{Proof}:
From Theorem 2 applied to the Hamiltonian of Eq.~\eqref{eq:changinghamiltonian} we get
\be
\nonumber
A_{\alpha}(\rho \otimes \ketbra{0}{0},H) \geq A_{\alpha}(\sigma \otimes \ketbra{1}{1},H).
\ee
But $
\mathcal{D}_H (\rho \otimes \ketbra{0}{0}) = \mathcal{D}_{H_0 \otimes \ketbra{0}{0}} (\rho \otimes \ketbra{0}{0}) = \mathcal{D}_{H_0} (\rho)\otimes \ketbra{0}{0}$. Hence $A_{\alpha}(\rho \otimes \ketbra{0}{0},H) = A_{\alpha}(\rho,H_0)$, and similarly for $H_1$. The result follows.

\subsection{Proof of Eq. (5)}

We show that for any qubit state $\rho$ and $\alpha \geq 0$, 
\be
0 \leq A_{\alpha}(\rho^{\otimes n}) \leq \log (n+1). 
\ee

\emph{Proof}: without loss of generality we can fix the Hamiltonian of the system to be the Pauli $Z$. Assume we are able to prove the result for every pure qubit state $\ket{\psi}$. Then for every $\rho$ there exists $p$ and $\ket{\psi}$ such that
\be
\rho = p \ketbra{\psi}{\psi} + (1-p) \I /2 := \mathcal{E}_{\mathrm{mix}}(\ketbra{\psi}{\psi})
\ee
Mixing with the identity is a time-translation symmetric operation:
\begin{eqnarray*}
\mathcal{E}_{\mathrm{mix}}(e^{-i H t} \sigma e^{i H t}) &=& p e^{-i H t} \sigma e^{i H t} + (1-p) \I/2   \\ &=& e^{-i H t} \mathcal{E}_{\mathrm{mix}}(\sigma) e^{i H t}
\end{eqnarray*}
Hence we can map $\ketbra{\psi}{\psi}^{\otimes n} \rightarrow \rho^{\otimes n}$ by means of symmetric operations. However it is easy to see that Theorem 2 holds, more generally, for any symmetric operation, so
\be
\label{eq:puretomix}
A_{\alpha}(\ketbra{\psi}{\psi}^{\otimes n}) \geq A_{\alpha}(\rho^{\otimes n}), \quad \forall \alpha \geq 0.
\ee
We conclude that we need to prove the bound only for pure states and from Eq. \eqref{eq:puretomix} the result will follow for any state. Because rotations about $Z$ are symmetric operations, we can assume $\ket{\psi}$ to lie on the $xz$ plane of the Bloch sphere:
\be
\rho = \ketbra{\psi}{\psi} = 
\left[
\begin{matrix} p & \sqrt{p(1-p)} \\ \sqrt{p(1-p)} & p \end{matrix}
\right]
\ee
We will use the notation $\mathcal{D}_{\sum_{i=1}^n Z_i} \equiv \mathcal{D}$. Expanding in the computational basis,
\be
\label{d}
\mathcal{D}(\rho^{\otimes n}) = \bigoplus_{h = 0}^n  p^{n-h} (1-p)^h  \ketbra{\I_h}{\I_h}, \quad \ket{\I_h} = \underbrace{(1,...,1)}_{\binom{n}{h} \text{ elements}}.
\ee
It is useful to introduce a vector $\ket{v_n}$ whose components are grouped in blocks as follows: $1$ component equal to $\sqrt{p^n}$, $\binom{n}{1}$ components equal to $\sqrt{p^{n-1}(1-p)}$, ..., $\binom{n}{h}$ components equal to $\sqrt{p^{n-h}(1-p)^h}$, ..., $1$ component equal to $\sqrt{(1-p)^n}$. Then we can compactly rewrite
\be
\rho^{\otimes n} = \ketbra{v_n}{v_n}
\ee
Define 
\be
\nonumber
P=
\mathcal{D}(\rho^{\otimes n})^{\frac{1-\alpha}{2\alpha}}\ketbra{v_n}{v_n}\mathcal{D}(\rho^{\otimes n})^{\frac{1-\alpha}{2\alpha}}.
\ee
From the definition of $A_\alpha$,
\be
\label{antimes}
A_{\alpha}(\rho^{\otimes n}) = \frac{1}{\alpha -1}\log \mathrm{Tr}[P^{\alpha}],
\ee

From $\braket{\I_h}{\I_h} = \binom{n}{h}$, for all $\alpha >0$ we have
\be
\nonumber
(\ketbra{\I_h}{\I_h})^\frac{1-\alpha}{2\alpha} = \binom{n}{h}^{\left(\frac{1-\alpha}{2\alpha} - 1\right)} \ketbra{\I_h}{\I_h},
\ee
so that from Eq.~\eqref{d},
\be
\nonumber
\mathcal{D}(\rho^{\otimes n}) ^{\frac{1-\alpha}{2\alpha}} = \bigoplus_{h = 0}^n  \binom{n}{h}^\frac{1-3\alpha}{2\alpha}[p^{n-h}(1-p)^h]^{\frac{1-\alpha}{2\alpha}}\ketbra{\I_h}{\I_h}.
\ee
Define $\ket{w_n} = \mathcal{D}(\rho^{\otimes n}) ^{\frac{1-\alpha}{2\alpha}} \ket{v_n}$. Then $P= \ketbra{w_n}{w_n}$ and
\be
\nonumber
\ket{w_n} = \bigoplus_{h=0}^n \binom{n}{h}^{\frac{1-3\alpha}{2\alpha}}[p^{n-h}(1-p)^h]^{\frac{1-\alpha}{2\alpha}}\ket{\I_h}\bra{\I_h}\ket{v_n}.
\ee
The vector $\ket{w_n}$ is also grouped in blocks $h=0,1,...,n$, each of $\binom{n}{h}$ equal elements. One of the elements of the $h$-block can be found as follows: the block of ones $\ketbra{\I_h}{\I_h}$ sums the elements in the $h$-block of $\ket{v_n}$ (which are $\binom{n}{h}$ and identical), getting $\binom{n}{h} \sqrt{p^{n-h}(1-p)^h}$). Adding the prefactors we see that the elements of the $h$-block of $\ket{w_n}$ look like: 
\be
\nonumber
\binom{n}{h}^{\frac{1-3\alpha}{2\alpha}} [p^{n-h}(1-p)^h]^{\frac{1-\alpha}{2\alpha}}\binom{n}{h}p^{\frac{n-h}{2}}(1-p)^\frac{h}{2}.
\ee
We conclude
\be
\nonumber
\bra{w_n}= \left(1,..., \underbrace{\binom{n}{h}^{\frac{1-\alpha}{2\alpha}} [p^{n-h}(1-p)^h]^{\frac{1}{2\alpha}}}_{\binom{n}{h} \textrm{\; elements}},...,1 \right).
\ee
Assume $\alpha \in \mathbb{N}$. Then
\be
\label{tracepalpha}
\mathrm{Tr}[P^{\alpha}] = (\braket{w_n}{w_n})^\alpha.
\ee
But
\be
\braket{w_n}{w_n} = \sum_{h=0}^n \binom{n}{h}^{1/\alpha}p^{\frac{n-h}{\alpha}}(1-p)^{\frac{h}{\alpha}}.
\ee
Combining this and Eq.~\eqref{tracepalpha} we obtain
\be
\mathrm{Tr}[P^{\alpha}] = ||x(n)||_{1/\alpha},
\ee
with
\be
\nonumber
x(n) := \left\{ \binom{n}{0}p^n, ...,\binom{n}{h}p^{n-h}(1-p)^h, ....,\binom{n}{n}(1-p)^n \right\},
\ee
and we used the usual definition of $\ell^p$-norm
\be
\nonumber
\label{pnorms}
||x||_{p} = \left(\sum_i |x_i|^p \right)^{\frac{1}{p}}.
\ee
Assume $\alpha >1$. The monotonicity of $\ell^p$-norms
\be
q > r > 0 \Longrightarrow ||x(n)||_{q} \leq ||x(n)||_{r}
\ee
implies
\be
\nonumber
||x(n)||_{1/\alpha} \geq ||x(n)||_1 = \sum_{h=0}^n \binom{n}{h} p^{n-h}(1-p)^{h}=1
\ee
Now, because
\be
\label{tracealpha3}
A_{\alpha}(\rho^{\otimes n})  = \frac{1}{\alpha -1}\log ||x(n)||_{1/\alpha}.
\ee
this means that for $\alpha>1$ any upper bound on $||x(n)||_{1/\alpha}$ gives an upper bound on $A_{\alpha}$. 

Fix $\alpha >1$. We can now use the following identity concerning $p$-norms (that follows from H\"older inequality): for all $p > r > 0$, if $y$ is a sequence of $k$ elements,
\be
\nonumber
||y||_r \leq k^{\left(\frac{1}{r}-\frac{1}{p}\right)} ||y||_p.
\ee
Choose $p=1$, $r=1/\alpha$:
\be
||x(n)||_{1/\alpha}\leq (n+1)^{\alpha-1}
\ee 
Hence, substituting in Eq.~\eqref{tracealpha3},
\be
\nonumber
A_{\alpha}(\rho^{\otimes n})  \leq \frac{1}{\alpha -1}\log \left[ (n+1)^{\alpha-1} \right] = \log (n+1),
\ee
for all integers $\alpha>1$. From the monotonicity in $\alpha$ of $A_{\alpha}$ \cite{muller2013quantum}, it is easy to see that this implies the result for every $\alpha \geq 0$, as required. The other inequality, $A_{\alpha} \geq 0$, follows immediately from the properties of $S_{\alpha}$.

\subsection{Work-locking}
\label{meth:work}

\emph{Proof of Theorem 3}
The state $\mathcal{D}_H (\rho)$ can be obtained from $\rho$ through dephasing in energy, which is easily shown to be a time-translation symmetric operation. Hence any work distribution that can be obtained from $\mathcal{D}_H (\rho)$ can be obtained from $\rho$ as well. Conversely, suppose it is possible to obtain from $\rho$ a work distribution $p(w)$ through a time-symmetric operation $\mathcal{E}$:
\be
\label{eq:worklocking}
\mathcal{E}(\rho \otimes \ketbra{0}{0}) = \sigma \otimes \sum_w  p(w)\ketbra{w}{w}.
\ee
If we apply $\mathcal{E}$ to $\mathcal{D}_H (\rho)$ it is easy to see from Eq. \eqref{edcommute} and \eqref{eq:worklocking}
\be
\mathcal{E}(\mathcal{D}_H(\rho) \otimes \ketbra{0}{0}) = \mathcal{D}_H(\sigma) \otimes \sum_w p(w)\ketbra{w}{w}.
\ee
Hence any work distribution that can be extracted from $\rho$ can be obtained from $\mathcal{D}_H (\rho)$ as well through the same $\mathcal{E}$.

\textbf{Acknowledgements:} ML would like to thank K. Korzekwa, J. Oppenheim, F. Mintert, T. Tufarelli and A. Milne for useful discussions. DJ is supported by the Royal Society. TR is supported by the Leverhulme Trust. ML is supported in part by EPSRC, COST Action MP1209 and Fondazione Angelo Della Riccia. 
\\ \textbf{Author contributions}
\\ ML, DJ, TR designed the research, ML performed the calculations/developed the results, ML and DJ wrote the paper.
\\ \textbf{Competing financial interests} \\ The authors declare no competing financial interests.

\end{document}